\documentclass[manuscript, screen]{acmart}
\AtBeginDocument{%
  }

\setcopyright{rightsretained}
\copyrightyear{2025}
\acmYear{2025}
\acmDOI{XXXXXXX.XXXXXXX}

\acmConference[CSCW '25]{CSCW 2025 Workshop on Exploring Resistance and Other Oppositional Responses to AI, Bergen, Norway}{October 18--22, 2025}{Bergen, Norway}
\acmBooktitle{CSCW 2025 Workshop on Exploring Resistance and Other Oppositional Responses to AI, October 18--22, 2025, Bergen, Norway}

\acmISBN{978-1-4503-XXXX-X/2018/06}

\begin{document}

\title{Examining Solidarity Against AI-Enabled Surveillance at the Intersection of Workplace and Carceral Realities}

\author{Morgan McErlean}
\email{mmcerle1@swarthmore.edu}
\orcid{0009-0004-8436-924X}
\affiliation{%
  \institution{Swarthmore College}
  \city{Swarthmore}
  \state{Pennsylvania}
  \country{USA}
}

\author{Cella M. Sum}
\email{csum@andrew.cmu.edu}
\affiliation{%
  \institution{Carnegie Mellon University}
  \city{Pittsburgh}
  \state{Pennsylvania}
  \country{USA}
}

\author{Sukrit Venkatagiri}
\email{sukrit@swarthmore.edu}
\orcid{0000-0002-3888-7693}
\affiliation{%
  \institution{Swarthmore College}
  \city{Swarthmore}
  \state{Pennsylvania}
  \country{USA}
}

\author{Sarah E. Fox}
\email{sarahfox@cmu.edu}
\orcid{0000-0002-7888-2598}
\affiliation{%
  \institution{Carnegie Mellon University}
  \city{Pittsburgh}
  \state{Pennsylvania}
  \country{USA}
}

\renewcommand{\shortauthors}{McErlean, Sum, Venkatagiri}

\begin{abstract}
  As panoptical, AI-driven surveillance becomes a norm, everyone is impacted. In a reality where all people fall victim to these technologies, establishing links and solidarity is essential to fighting back. Two groups facing rising and targeted surveillance are workers and individuals impacted by the carceral system. Through preliminary data collection from a worker-surveillance lens, our findings reveal several cases of these surveillance infrastructures intersecting. Continuation of our work will involve collecting cases from a carceral-centered lens. Driven by a community-facing analysis of the overlap in the AI-driven surveillance experienced by workers and individuals impacted by the carceral system, we will facilitate discussions with restorative justice activists around cultivating solidarity and empowerment focused on the interconnected nature of workplace and carceral surveillance technologies.
\end{abstract}





\maketitle

\section{Introduction}
Surveillance is pervasive in our society from public spaces–on the street, in every store we enter–to “private” spaces such as offices, classrooms, and even, in some cases, our homes \cite{lyon, marx, owens22, zuboff}.  The increased integration of AI into surveillance systems enables cameras armed with facial recognition technologies, monitoring algorithms to detect individuals' behaviors, and biometric systems that attempt to predict physiological states \cite {agre, awumey, scheuerman, stark}. Prior work in HCI and CSCW has explored surveillance, noting its reliance on technological and algorithmic systems, from a variety of lenses. These works demonstrate the use of AI/algorithms in surveillance is pervasive in both public and private spaces. Sex workers are subjected to facial recognition systems that block their use of certain social platforms \cite{hamilton}. Workers across all sectors, remote and in-person, have been subjected to electronic monitoring that relies on algorithms to assess their productivity, efficiency, and even feelings, whether it be through metrics determined by audio/visual recordings, screen tracking, activity tracking, or collection of biometric data \cite{awumey, sum, zickhur}. Carceral surveillance systems are another area which has solicited some attention. Works examine communication with people inside prison, electronic monitoring apps for people on probation and parole, and the rise of a state supervised society \cite{owens22, owens21, miller}. In alignment with these previous works, our work examines two communities facing targeted AI-enabled surveillance: workers and individuals impacted by the carceral system\footnote{ For purposes of this work, individuals impacted by the carceral system will refer to people who are currently in prison, currently under some form of state supervision (parolees, people on probation), were incarcerated or under some form of supervision at any point in their lives, or have otherwise been impacted by the carceral system (have been arrested, have a mugshot, have been to jail).
}. Through such an examination, our work attempts to link these surveillance architectures and through establishing this link, search for ways to foster solidarity in the face of a proliferation of AI-enabled surveillance.

\section{Examining Intersections Between Workplace and Carceral Surveillance}
Nowhere is surveillance more apparent than in the carceral system, but, as demonstrated by previous works, other populations, including workers, face privacy violating surveillance practices that result in a decrease in autonomy and increasing exploitation
\cite{owens21, calacci, sum}. It is essential to note, this connection being made is \textbf{not} an attempt to conflate these veillance architectures. People victimized by the prison system are forced to knowingly comply with the privacy and autonomy violations they face and are often seen as deserving invasive surveillance \cite{owens21, garvie}; workers face overlapping surveillance practices, but overall, workers have autonomy to escape these practices (e.g quitting) \cite{sum}. In fact this very perception, of people impacted by the carceral system being deserving of this surveillance, drove the authors’ interest in a project where cross-movement solidarity could be possible to build. A major hurdle in carceral research, more generally, and AI-enabled surveillance examinations of carceral realities specifically, is a conception of these individuals deserving the treatment they receive on the part of the state and society at large, whether that be ostracization or other forms of punishment. Distorting this perception is a formidable task.

Predictably, carceral realities have received only occasional attention from the HCI community, despite AI-enabled surveillance of this community being pervasive and evergrowing \cite{miller, owens22, owens21, ogburu, gautam}. Facial recognition technologies (FRTs) have enabled law enforcement to track and criminalize carceral system victims due to many FRTs’ reliance on mugshot databases and FRTs’ inaccuracy in identifying people of color \cite{zuboff, buolamwini, owens24, prison, wessler}. COMPAS\footnote{Correctional Offender Management Profiling for Alternative Sanctions, a case management system aimed at predicting the likelihood of a defendant becoming a recidivist.} is another example of an algorithmic system whose racial biases impacted returning citizens; it is an entire architecture of these systems that contributes to the landscape that returning citizens must face, dubbed “carceral citizenship” by Miller et al \cite{zickhur, miller}. Further, Sugie states that returning citizens often rely on work that is irregular and temporary “gigs” \cite{sugie}; gig workers face black box algorithmic systems deployed by the apps that they work for, which heavily control, surveil, and dictate their work \cite{shannon, irani}. 

Beyond an overlap that exists in heavily surveilled work and work performed by many returning citizens, in what ways do these systems themselves cross both workplace and carceral spaces? Crucially absent from the literature is a cross-examination of the surveillance that exists within workplaces and carceral systems, specifically technological systems used in both worker and carceral spaces and dual purpose surveillance architectures (e.g. surveilling workers and trying to deter crime in a workplace). Assessing these crossovers and distilling it in a public facing way can present a conduit by which to engage in discussions surrounding surveillance realities shared in carceral spaces and the workplace aimed at fostering solidarity and resistance against AI-enabled surveillance. 

\section{Preliminary Findings}

In preliminary findings, we have discovered several examples of cross-pollination between these two surveillance spaces. These cases are from a database of workplace surveillance incidents collected from journalistic sources for a project at CMU’s Tech Solidarity Lab, currently in the works with PhD candidate Cella M. Sum, advised by Dr. Sarah Fox. A crossover in these surveillance systems has been demonstrated by Rite-Aid’s use of FRTs to surveil workers and deter crime at locations in predominately black and low-income neighborhoods \cite{dastin}; Amazon implying they have access to Thomson Reuters CLEAR, an AI-powered tool typically only licensed to law enforcement \cite{negron}; and workplace productivity monitoring software, Veriato, being used to monitor people on parole \cite{karageorgos}. These findings encompass an examination from the lens of worker-surveillance; further work on this project will involve more extensive data collection from a carceral-surveillance lens.

In the second phase of this project, we will conduct group discussions with restorative justice activists \footnote{ Restorative justice is a framework for imagining healing after returning from incarceration. As articulated by Angela Davis, it means “to imagine a society that is secure,” without prisons \cite{davis}.} to consider alternatives and actions against the proliferation of these AI-based surveillance technologies. Thus far, informal conversations with community partners at abolitionist and restorative justice organizations in Philadelphia have highlighted concerns about "spending money on technology we know is racist and functioning in that way” and an “illusion of freedom while you are actually being watched.” This abolitionist perspective serves a reminder of a hope for a future unburdened by these oppressive systems. Through this examination of surveillance intersections that exist within these communities, we hope to parse cross-movement (worker, carceral) solidarity and overlap in the struggle with regard to surveillance technologies, situated in and proliferated by a growing AI space, hoping to foster solidarity in resistance against these systems.

\begin{acks}
Thank you to Cella Sum, Dr. Sarah Fox, and the Tech Solidarity Lab at CMU HCII for allowing me to extend off an ongoing project for purposes of this work. 
\end{acks}

\bibliographystyle{ACM-Reference-Format}
\bibliography{bib}


\begin{thebibliography}{28}


\ifx \showCODEN    \undefined \def \showCODEN     #1{\unskip}     \fi
\ifx \showISBNx    \undefined \def \showISBNx     #1{\unskip}     \fi
\ifx \showISBNxiii \undefined \def \showISBNxiii  #1{\unskip}     \fi
\ifx \showISSN     \undefined \def \showISSN      #1{\unskip}     \fi
\ifx \showLCCN     \undefined \def \showLCCN      #1{\unskip}     \fi
\ifx \shownote     \undefined \def \shownote      #1{#1}          \fi
\ifx \showarticletitle \undefined \def \showarticletitle #1{#1}   \fi
\ifx \showURL      \undefined \def \showURL       {\relax}        \fi
\providecommand\bibfield[2]{#2}
\providecommand\bibinfo[2]{#2}
\providecommand\natexlab[1]{#1}
\providecommand\showeprint[2][]{arXiv:#2}

\bibitem[Agre(1994)]%
        {agre}
\bibfield{author}{\bibinfo{person}{Philip~E. Agre}.} \bibinfo{year}{1994}\natexlab{}.
\newblock \showarticletitle{Surveillance and capture: Two models of privacy}.
\newblock \bibinfo{journal}{\emph{The Information Society}} \bibinfo{volume}{10}, \bibinfo{number}{2} (\bibinfo{date}{April} \bibinfo{year}{1994}), \bibinfo{pages}{101–127}.
\newblock
\href{https://doi.org/10.1080/01972243.1994.9960162}{doi:\nolinkurl{10.1080/01972243.1994.9960162}}


\bibitem[Awumey et~al\mbox{.}(2024)]%
        {awumey}
\bibfield{author}{\bibinfo{person}{Ezra Awumey}, \bibinfo{person}{Sauvik Das}, {and} \bibinfo{person}{Jodi Forlizzi}.} \bibinfo{year}{2024}\natexlab{}.
\newblock \showarticletitle{A Systematic Review of Biometric Monitoring in the Workplace: Analyzing Socio-technical Harms in Development, Deployment and Use}. In \bibinfo{booktitle}{\emph{The 2024 ACM Conference on Fairness, Accountability, and Transparency, Rio de Janeiro Brazil: ACM}}. \bibinfo{publisher}{ACM Press}, \bibinfo{pages}{920–932}.
\newblock
\href{https://doi.org/10.1145/3630106.3658945}{doi:\nolinkurl{10.1145/3630106.3658945}}


\bibitem[Buolamwini(2017)]%
        {buolamwini}
\bibfield{author}{\bibinfo{person}{Joy~A. Buolamwini}.} \bibinfo{year}{2017}\natexlab{}.
\newblock \emph{\bibinfo{title}{Gender shades: intersectional phenotypic and demographic evaluation of face datasets and gender classifiers}}.
\newblock \bibinfo{thesistype}{Ph.\,D. Dissertation}. \bibinfo{school}{Massachusetts Institute of Technology}, \bibinfo{address}{Cambridge, MA}.
\newblock
\urldef\tempurl%
\url{https://dspace.mit.edu/handle/1721.1/114068}
\showURL{%
\tempurl}


\bibitem[Calacci(2022)]%
        {calacci}
\bibfield{author}{\bibinfo{person}{Dan Calacci}.} \bibinfo{year}{2022}\natexlab{}.
\newblock \showarticletitle{Organizing in the End of Employment: Information Sharing, Data Stewardship, and Digital Workerism}. In \bibinfo{booktitle}{\emph{2022 Symposium on Human-Computer Interaction for Work, Durham NH USA: ACM}}. \bibinfo{publisher}{ACM Press}, \bibinfo{pages}{1–9}.
\newblock
\href{https://doi.org/10.1145/3533406.353342}{doi:\nolinkurl{10.1145/3533406.353342}}


\bibitem[Dastin(2020)]%
        {dastin}
\bibfield{author}{\bibinfo{person}{Jeffery Dastin}.} \bibinfo{year}{2020}\natexlab{}.
\newblock \bibinfo{booktitle}{\emph{Special Report: Rite Aid deployed facial recognition systems in hundreds of U.S. stores}}.
\newblock
\urldef\tempurl%
\url{https://www.reuters.com/article/technology/special-report-rite-aid-deployed-facial-recognition-systems-in-hundreds-of-us-idUSKCN24T1FT}
\showURL{%
\tempurl}


\bibitem[Davis(2016)]%
        {davis}
\bibfield{author}{\bibinfo{person}{Angela Davis}.} \bibinfo{year}{2016}\natexlab{}.
\newblock \bibinfo{booktitle}{\emph{Freedom is a Constant Struggle: Ferguson, Palestine, and the Foundations of a Movement}}.
\newblock \bibinfo{publisher}{Haymarket Books}.
\newblock


\bibitem[Garvie et~al\mbox{.}(2016)]%
        {garvie}
\bibfield{author}{\bibinfo{person}{Clare Garvie}, \bibinfo{person}{Alvaro Bedoya}, {and} \bibinfo{person}{Jonathan Frankle}.} \bibinfo{year}{2016}\natexlab{}.
\newblock \bibinfo{booktitle}{\emph{The Perpetual Line-Up}}.
\newblock
\urldef\tempurl%
\url{https://www.perpetuallineup.org/}
\showURL{%
\tempurl}


\bibitem[Gautam et~al\mbox{.}(2024)]%
        {gautam}
\bibfield{author}{\bibinfo{person}{Aakash Gautam}, \bibinfo{person}{Khushboo Gandhi}, {and} \bibinfo{person}{Jessica Sendejo}.} \bibinfo{year}{2024}\natexlab{}.
\newblock \showarticletitle{Enhancing Reentry Support Programs Through Digital Literacy Integration}. In \bibinfo{booktitle}{\emph{DIS '24: Proceedings of the 2024 ACM Designing Interactive Systems Conference}}. \bibinfo{pages}{2882 -- 2896}.
\newblock
\href{https://doi.org/10.1145/3643834.3660730}{doi:\nolinkurl{10.1145/3643834.3660730}}


\bibitem[Hamilton et~al\mbox{.}(2024)]%
        {hamilton}
\bibfield{author}{\bibinfo{person}{Vaughn Hamilton}, \bibinfo{person}{Gabriel Kaptchuk}, \bibinfo{person}{Allison McDonald}, {and} \bibinfo{person}{Elissa~M. Redmiles}.} \bibinfo{year}{2024}\natexlab{}.
\newblock \bibinfo{booktitle}{\emph{Safer Digital Intimacy For Sex Workers And Beyond: A Technical Research Agenda}}.
\newblock
\href{https://doi.org/10.48550/arXiv.2403.10688}{doi:\nolinkurl{10.48550/arXiv.2403.10688}}


\bibitem[Initiative({[n.\,d.]})]%
        {prison}
\bibfield{author}{\bibinfo{person}{Prison~Policy Initiative}.} \bibinfo{year}{[n.\,d.]}\natexlab{}.
\newblock \bibinfo{booktitle}{\emph{Beyond the Count: A deep dive into state prison populations.}}
\newblock
\urldef\tempurl%
\url{https://www.prisonpolicy.org/reports/beyondthecount.html}
\showURL{%
\tempurl}


\bibitem[Irani and Silberman(2013)]%
        {irani}
\bibfield{author}{\bibinfo{person}{Lilly~C. Irani} {and} \bibinfo{person}{M.~Six Silberman}.} \bibinfo{year}{2013}\natexlab{}.
\newblock \showarticletitle{Turkopticon: interrupting worker invisibility in amazon mechanical turk}. In \bibinfo{booktitle}{\emph{Proceedings of the SIGCHI Conference on Human Factors in Computing Systems, Paris France: ACM}}. \bibinfo{publisher}{ACM Press}, \bibinfo{pages}{611–620}.
\newblock
\href{https://doi.org/10.1145/2470654.2470742}{doi:\nolinkurl{10.1145/2470654.2470742}}


\bibitem[Karageorgos(2023)]%
        {karageorgos}
\bibfield{author}{\bibinfo{person}{Panos Karageorgos}.} \bibinfo{year}{2023}\natexlab{}.
\newblock \bibinfo{booktitle}{\emph{Fighting Recidivism of High-Risk Offenders With Veriato}}.
\newblock
\urldef\tempurl%
\url{https://veriato.com/case-studies/fighting-recidivism-of-high-risk-offenders-with-veriato/}
\showURL{%
\tempurl}


\bibitem[Lyon(2010)]%
        {lyon}
\bibfield{author}{\bibinfo{person}{David Lyon}.} \bibinfo{year}{2010}\natexlab{}.
\newblock \showarticletitle{Surveillance, power and everyday life}.
\newblock \bibinfo{journal}{\emph{Emerging digital spaces in contemporary society: Properties of technology (2010)}} (\bibinfo{year}{2010}), \bibinfo{pages}{107–120}.
\newblock


\bibitem[Marx(2002)]%
        {marx}
\bibfield{author}{\bibinfo{person}{Gary~T. Marx}.} \bibinfo{year}{2002}\natexlab{}.
\newblock \showarticletitle{What’s New About the ‘New Surveillance’? Classifying for Change and Continuity.}
\newblock \bibinfo{journal}{\emph{Surveillance and Society}} \bibinfo{volume}{1}, \bibinfo{number}{1} (\bibinfo{year}{2002}), \bibinfo{pages}{9–29}.
\newblock
\href{https://doi.org/10.24908/ss.v1i1.3391}{doi:\nolinkurl{10.24908/ss.v1i1.3391}}


\bibitem[Miller and Alexander(2016)]%
        {miller}
\bibfield{author}{\bibinfo{person}{Reuben~Jonathan Miller} {and} \bibinfo{person}{Amanda Alexander}.} \bibinfo{year}{2016}\natexlab{}.
\newblock \showarticletitle{The Price of Carceral Citizenship: Punishment, Surveillance, and Social Welfare Policy in an Age of Carceral Expansion}.
\newblock \bibinfo{journal}{\emph{The Michigan Journal of Race and Law}} \bibinfo{volume}{21}, \bibinfo{number}{2} (\bibinfo{year}{2016}), \bibinfo{pages}{291}.
\newblock
\href{https://doi.org/10.36643/mjrl.21.2.price}{doi:\nolinkurl{10.36643/mjrl.21.2.price}}


\bibitem[Negrón(2021)]%
        {negron}
\bibfield{author}{\bibinfo{person}{Wilneida Negrón}.} \bibinfo{year}{2021}\natexlab{}.
\newblock \bibinfo{booktitle}{\emph{Little Tech Is Coming for Workers}}.
\newblock
\urldef\tempurl%
\url{https://home.coworker.org/wp-content/uploads/2021/11/Little-Tech-Is-Coming-for-Workers.pdf}
\showURL{%
\tempurl}


\bibitem[Ogbonnaya-Ogburu et~al\mbox{.}(2025)]%
        {ogburu}
\bibfield{author}{\bibinfo{person}{Ihudiya~Finda Ogbonnaya-Ogburu}, \bibinfo{person}{Edward~(Barakah) Sanders}, \bibinfo{person}{Johnny Alexander-Bey}, \bibinfo{person}{Brandon Harrington}, \bibinfo{person}{Myron Wood}, \bibinfo{person}{Tawanna~R. Dillahunt}, {and} \bibinfo{person}{Kentaro Toyama}.} \bibinfo{year}{2025}\natexlab{}.
\newblock \showarticletitle{Designing Digital Tools to Support Online Job Search for Returning Citizens}. In \bibinfo{booktitle}{\emph{COMPASS '25: Proceedings of the ACM SIGCAS/SIGCHI Conference on Computing and Sustainable Societies}}. \bibinfo{pages}{339 -- 353}.
\newblock
\href{https://doi.org/10.1145/3715335.3735475}{doi:\nolinkurl{10.1145/3715335.3735475}}


\bibitem[Owens et~al\mbox{.}(2022)]%
        {owens22}
\bibfield{author}{\bibinfo{person}{Kentrell Owens}, \bibinfo{person}{Anita Alem}, \bibinfo{person}{Franziska Roesner}, {and} \bibinfo{person}{Tadayoshi Kohno}.} \bibinfo{year}{2022}\natexlab{}.
\newblock \showarticletitle{Electronic Monitoring Smartphone Apps: An Analysis of Risks from Technical, Human-Centered, and Legal Perspectives}. In \bibinfo{booktitle}{\emph{USENIX Security, 2022}}.
\newblock
\urldef\tempurl%
\url{https://www.usenix.org/conference/usenixsecurity22/presentation/owens}
\showURL{%
\tempurl}


\bibitem[Owens et~al\mbox{.}(2021)]%
        {owens21}
\bibfield{author}{\bibinfo{person}{Kentrell Owens}, \bibinfo{person}{Camille Cobb}, {and} \bibinfo{person}{Lorrie Cranor}.} \bibinfo{year}{2021}\natexlab{}.
\newblock \showarticletitle{You Gotta Watch What You Say’: Surveillance of Communication with Incarcerated People}. In \bibinfo{booktitle}{\emph{Proceedings of the 2021 CHI Conference on Human Factors in Computing Systems, Yokohama Japan: ACM}}. \bibinfo{pages}{1--8}.
\newblock
\href{https://doi.org/10.1145/3411764.3445055}{doi:\nolinkurl{10.1145/3411764.3445055}}


\bibitem[Owens et~al\mbox{.}(2024)]%
        {owens24}
\bibfield{author}{\bibinfo{person}{Kentrell Owens}, \bibinfo{person}{Camille Cobb}, {and} \bibinfo{person}{Lorrie Cranor}.} \bibinfo{year}{2024}\natexlab{}.
\newblock \showarticletitle{Face the Facts: Using Face Averaging to Visualize Gender-by-Race Bias in Facial Analysis Algorithms}. In \bibinfo{booktitle}{\emph{AIES}}. \bibinfo{pages}{1101–1111}.
\newblock
\href{https://doi.org/10.1609/aies.v7i1.31707}{doi:\nolinkurl{10.1609/aies.v7i1.31707}}


\bibitem[Sannon et~al\mbox{.}(2022)]%
        {shannon}
\bibfield{author}{\bibinfo{person}{Shruti Sannon}, \bibinfo{person}{Billie Sun}, {and} \bibinfo{person}{Dan Cosley}.} \bibinfo{year}{2022}\natexlab{}.
\newblock \showarticletitle{Privacy, Surveillance, and Power in the Gig Economy}. In \bibinfo{booktitle}{\emph{CHI Conference on Human Factors in Computing Systems, New Orleans LA USA: ACM}}. \bibinfo{pages}{1–15}.
\newblock
\href{https://doi.org/10.1145/3491102.3502083}{doi:\nolinkurl{10.1145/3491102.3502083}}


\bibitem[Scheuerman et~al\mbox{.}(2021)]%
        {scheuerman}
\bibfield{author}{\bibinfo{person}{Morgan~Klaus Scheuerman}, \bibinfo{person}{Alex Hanna}, {and} \bibinfo{person}{Remi Denton}.} \bibinfo{year}{2021}\natexlab{}.
\newblock \showarticletitle{Do Datasets Have Politics? Disciplinary Values in Computer Vision Dataset Development}. In \bibinfo{booktitle}{\emph{Proc. ACM Hum.-Comput. Interact., vol. 5, no. CSCW2}}. \bibinfo{pages}{1–37}.
\newblock
\href{https://doi.org/10.1145/3476058}{doi:\nolinkurl{10.1145/3476058}}


\bibitem[Stark(2019)]%
        {stark}
\bibfield{author}{\bibinfo{person}{Luke Stark}.} \bibinfo{year}{2019}\natexlab{}.
\newblock \showarticletitle{Facial recognition is the plutonium of AI}.
\newblock \bibinfo{journal}{\emph{XRDS}} \bibinfo{number}{3} (\bibinfo{date}{April} \bibinfo{year}{2019}), \bibinfo{pages}{50--55}.
\newblock
\href{https://doi.org/10.1145/3313129}{doi:\nolinkurl{10.1145/3313129}}


\bibitem[Sugie(2018)]%
        {sugie}
\bibfield{author}{\bibinfo{person}{Naomi~F. Sugie}.} \bibinfo{year}{2018}\natexlab{}.
\newblock \showarticletitle{Work as Foraging: A Smartphone Study of Job Search and Employment after Prison}.
\newblock \bibinfo{journal}{\emph{Amer. J. Sociology}} \bibinfo{number}{5} (\bibinfo{date}{March} \bibinfo{year}{2018}), \bibinfo{pages}{1453–1491}.
\newblock
\href{https://doi.org/10.1086/696209}{doi:\nolinkurl{10.1086/696209}}


\bibitem[Sum et~al\mbox{.}(2025)]%
        {sum}
\bibfield{author}{\bibinfo{person}{Cella~M. Sum}, \bibinfo{person}{Caroline Shi}, {and} \bibinfo{person}{Sarah~E. Fox}.} \bibinfo{year}{2025}\natexlab{}.
\newblock \showarticletitle{It’s Always a Losing Game’: How Workers Understand and Resist Surveillance Technologies on the Job}. In \bibinfo{booktitle}{\emph{Proc. ACM Hum.-Comput. Interact.}} \bibinfo{pages}{1–32}.
\newblock
\href{https://doi.org/10.1145/3710902}{doi:\nolinkurl{10.1145/3710902}}


\bibitem[Wessler(2024)]%
        {wessler}
\bibfield{author}{\bibinfo{person}{Nathan~Freed Wessler}.} \bibinfo{year}{2024}\natexlab{}.
\newblock \bibinfo{booktitle}{\emph{Police Say a Simple Warning Will Prevent Face Recognition Wrongful Arrests. That's Just Not True.}}
\newblock
\urldef\tempurl%
\url{https://www.aclu.org/news/privacy-technology/police-say-a-simple-warning-will-prevent-face-recognition-wrongful-arrests-thats-just-not-true}
\showURL{%
\tempurl}


\bibitem[Zickuhr(2021)]%
        {zickhur}
\bibfield{author}{\bibinfo{person}{Kathryn Zickuhr}.} \bibinfo{year}{2021}\natexlab{}.
\newblock \bibinfo{booktitle}{\emph{Workplace surveillance is becoming the new normal for U.S. workers}}.
\newblock
\urldef\tempurl%
\url{https://equitablegrowth.org/research-paper/workplace-surveillance-is-becoming-the-new-normal-for-u-s-workers/}
\showURL{%
\tempurl}


\bibitem[Zuboff(2019)]%
        {zuboff}
\bibfield{author}{\bibinfo{person}{Shoshana Zuboff}.} \bibinfo{year}{2019}\natexlab{}.
\newblock \bibinfo{booktitle}{\emph{The Age of Surveillance Capitalism: The Fight for a Human Future at the New Frontier of Power - Book - Faculty and Research - Harvard Business School}}.
\newblock
\urldef\tempurl%
\url{https://www.hbs.edu/faculty/Pages/item.aspx?num=56791}
\showURL{%
\tempurl}


\end{thebibliography}

\end{document}